\documentclass{article} 
\usepackage{psfig} 
\usepackage{a4wide}  
\usepackage{verbatim} 
 
\begin{document} 

\title{Coherent multiple scattering in disordered media}

\author{E.~Akkermans$^1$ and G.~Montambaux$^2$  
 \\$^1$Department of Physics,  
 Technion, 32000 Haifa, Israel \\  $^2$Laboratoire de Physique des Solides
\\Universit\'e Paris-Sud
\\Bat.510, F-91405 Orsay Cedex, France}

\maketitle

\begin{abstract}
These notes contain a rapid overview of the methods and results obtained
in the field of propagation of waves in disordered media. The case of
Schr\"odinger and Helmholtz equations are considered that describe respectively
electrons in metals and scalar electromagnetic waves. The assumptions on the nature
of disorder are discussed and perturbation methods in the weak disorder limit are presented.
A central quantity, namely the probability of quantum diffusion is defined and calculated
in the same limit. It is then shown that several relevant physical quantities are related
to the return probability. Examples are provided to substantiate this, which include
the average electrical conductivity, its fluctuations, the average albedo and spectral correlations.
\end{abstract}

\section {Introduction}

The study of wave propagation in random media gave rise to a huge
amount of work especially during the last twenty years. Today,
this field is split in two main subfields. One is concerned with
the interplay between coherence and disorder in metallic systems
and the second deals with the same problematics but for
electromagnetic waves (this includes as well sound or gravity
waves). Each one of the subfields has its own specificities and
advantages,
 such as the effect
of magnetic fields on transport and thermodynamics, interactions
for metals and
 angular structure (spectroscopy) for waves. Unfortunately, the split between
these two subfields has been growing so that their interplay is
now certainly too weak in spite of the existence of a number of
excellent reviews. These notes represent a very preliminary step
towards a unified presentation of this field \cite{am}. We have
tried to present a general formalism that can apply to both
situations (electrons and waves). It is centered around the
existence of a basic quantity {\it the probability of quantum
diffusion} which allows to describe either weak localization
effects of the electrical conductivity of metals, spectral
quantities of isolated electronic systems or the coherent and
incoherent albedo, dynamical effects in multiple scattering of
light by suspensions etc.

The next three sections contain basic definitions and generalities on the kind of waves we consider
and the model of disorder. The section 5 deals with the definition and the calculation of the
probability of quantum diffusion in multiple scattering using various approximations.
Then, the subsequent sections apply these results to a selection of examples taken either from
electronic systems (conductivity and spectral correlations)
or multiple scattering of electromagnetic waves (average albedo in optical systems).

\section{Models for the disordered potential}

We shall consider mainly two  problems where waves and disorder are involved. The
first one corresponds to the study of spinless electrons in disordered metals or semiconductors.
For a degenerate gas of free electrons of mass $m$ and charge $-e$, the Schr\"odinger
equation is governed by the Hamiltonian
\begin{equation}
{\cal H} \psi ({\bf r}) =  - {{\hbar^2} \over {2m}}
( {\bf \nabla} + {{ie} \over \hbar} {\bf A}{)^2}
 \psi ({\bf r})
+ V({ \bf r}) \psi ({\bf r}) \ \ .
\label{hamiltonien}
\end{equation}
where ${\bf B} =
{\bf \nabla} \times {\bf A}$ is the magnetic field. Effects associated to the band structure or
to interactions between the electrons in the framework of Fermi liquid theory are accounted for
by the replacement of the mass $m$ by an effective mass. The potential $V( {\bf r})$ decribes
both the scattering by inhomogeneities and the confinement potential.

The second problem we shall consider is the propagation of electromagnetic waves of frequency
$\omega$ in the {\it scalar}
approximation. The behaviour of the electric field $\psi ({\bf r})$
is obtained from the Helmholtz equation
\begin{equation}
- \Delta \psi ({\bf r}) - {k_0^2} \mu ({\bf r})  \psi ({\bf r}) =  {k_0^2}
 \psi ({\bf r})
\label{helmholtz}
\end{equation}
where the function $\mu ({\bf r}) = { \delta \epsilon / \overline \epsilon}$ is the
relative fluctuation of the dielectric constant,
 $k_0 = {\overline n} {{\omega} \over {c}}$ and ${\overline n}$ is the average optical index.
Under this form, the Helmholtz equation has a structure similar to the Schr\"odinger
equation and the waves are scattered by the fluctuations of the dielectric function. It is
nevertheless interesting to notice that in the latter case, the strength $\mu$ of the potential
is multiplied by the frequency $\omega^2$ so that in contrast with electronic systems, a decrease of
the frequency $\omega$ leads to a weaker effect of the disorder.

To account for the effects of the disorder in either case, we shall consider a random
continuous function
$V ({\bf r})$, of zero spatial average, $\langle V({\bf r}) \rangle =0 $, where $\langle \cdots \rangle$
represents the spatial disorder  average. The potential is characterized by its correlations
especially by the two-point correlation function
\begin{equation}
\langle V({\bf {r}}) V({\bf {r'}}) \rangle = B({\bf r} - {\bf r}')
\label{corrdesordre}
\end{equation}
For a potential $V({\bf {r}})$ which is localized enough, there
exits a length $r_c$ which describes the fall off of $B({\bf r} -
{\bf r}')$. In the limit where the wavelength $\lambda \gg r_c$,
the scattering events are statistically independent and we can
consider the limiting case
\begin{equation}
\langle V({\bf {r}}) V({\bf {r'}}) \rangle = \gamma \delta ({\bf
r} - {\bf r}') \label{bb}
\end{equation}
Such a potential is usually called a white noise. This is
the case we shall consider throughout these notes. For the case of the Helmholtz equation, the potential 
is taken to be $ V({\bf {r}}) = {k_0^2} \mu ({\bf r})$ so that $\gamma$, in that case, has the dimensions 
of the inverse of a length.

\section{Perturbation theory for multiple scattering}

\subsection{Single scattering - elastic scattering time}

For a very dilute system, we can first assume that a given incident wave of wavevector $\bf k$
is scattered only once into a state $\bf k'$ before leaving the system.
The lifetime $\tau_k$ of the state $\bf k$ is then given (using the notations of
quantum mechanics) by the Fermi golden rule
\begin{equation}
{1 \over \tau_{{\bf k}}} = {2 \pi \over \hbar} \sum_{{\bf k}'}
|\langle {\bf k} | V | {\bf k}' \rangle|^2 \delta(\epsilon_{\bf
k} - \epsilon_{{\bf k}'}) \ \ .
\end{equation}
For the white noise case, we obtain
\begin{equation}
{1 \over \tau_e} = {2 \pi \over \hbar} \rho_0 \gamma
\end{equation}
where $\rho_0$ is the density of states per unit volume.
To this time, we can associate a length, the elastic mean free path $l_e$, defined as
$l_e = v \tau_e$ by using the group velocity $v$.
But the Fermi golden rule is valid for
short times $t \ll \tau_e$ and in perturbation with the scattering potential $V$.
In order to go beyond these limitations, we need to resort to a more powerful tool namely the
formalism of the Green functions.

The Green function associated to a wave equation can be defined
as the response to a pulse i.e. to a $\delta$-function
perturbation. For the case of the hamiltonian
(\ref{hamiltonien}), the Green function $G^{R,A}({\bf r}_i,{\bf
r},\epsilon)$ is solution of
\begin{equation}
\left( \epsilon - {\cal H} \pm i0 \right) G^{R,A}({\bf r}_i, {\bf
r}, \epsilon) = \delta ({\bf r} - {\bf r}_i) \label{green2}
\end{equation}
We can define as well the free Green
function $G_0({\bf r}_i,{\bf r},\epsilon)$ in the absence of
scattering potential, which is the solution of
\begin{equation}
\left( \epsilon + {\hbar^2 \over 2 m} \Delta_{\bf r}  \pm i0
\right) G_0^{R,A}({\bf r}_i, {\bf r}, \epsilon) = \delta ({\bf r}
- {\bf r}_i) \label{green3}
\end{equation}
so that the Green function $G^{R,A}({\bf r}_i,{\bf r},\epsilon)$
can be also expressed as a solution of the integral equation
\begin{equation} G({\bf r}_i,{\bf r},\epsilon)=G_0({\bf r}_i,{\bf
r},\epsilon)+\int G({\bf r}_i,{\bf r}',\epsilon)V({\bf
r}')G_0({\bf r}',{\bf r},\epsilon) d{\bf r}' \label{integralG}
\end{equation} the solution of (\ref{green3}) is given (for d=3) by
\begin{equation}
G_0^{R,A}({\bf r}_i,{\bf r}, \epsilon)  =   - {m  \over {2 \pi }
\hbar^2} { e^{\pm i k R} \over  R} \label{green3d1}
\end{equation}
with $\epsilon = \hbar^2 k^2 / 2m$.

For the Helmholtz equation (\ref{helmholtz}), we obtain a
sequence of similar equations, but attention needs to be paid to
the fact that the dispersion of the waves is now linear instead
of quadratic for the Schr\"odinger case. The Green equation is
\begin{equation} \left( \Delta_{{\bf r}} + k_0^2 (1 + \mu ({\bf
r})) \right) G ({\bf r}_i, {\bf r},k_0) = \delta ({\bf r} - {\bf
r}') \label{greenj} \end{equation} while the solution of the free
Green equation obtained for $\mu ({\bf r}) = 0$, is
\begin{equation}
G^{R,A}_0 ({\bf r}_i,{\bf r},k_0)= - {1\over 4 \pi} {e^{\pm ik_0R}
\over  R} \label{green5}
\end{equation}

\subsection{Electromagnetic waves}

The formalism of Green functions provides an appropriate technical framework for the study
of solutions of wave equations for free systems namely without sources.
 It becomes essential for the study of the
propagation of electromagnetic waves from a distribution of
sources $j ({\bf r})$ of the field. This is the problem of
radiative transfer \cite{chandra}. For a pointlike source,
we are back to the previous problem of the Green equation. In the
general case, The Helmholtz equation (\ref{helmholtz})  needs to
be replaced by
\begin{equation}
 \Delta \psi ({\bf r}) + {k_0^2}(1 + \mu ({\bf r}))  \psi ({\bf r}) =
j({\bf r})
\label{radiatif}
\end{equation}
where $\psi ({\bf r})$ is indeed a Green function i.e. it depends
on the distribution of sources $j ({\bf r})$. The equation
(\ref{radiatif}) can also be written in the form of the integral
equation
\begin{equation}
\psi({\bf r}) = \int d {\bf r}_i  j({\bf r}_i) G_0 ({\bf r}_i,
{\bf r},k_0) - k_0^2 \int d{\bf r}' \psi({\bf r}')\mu ({\bf r}')
G_0({\bf r}', {\bf r},k_0) \label{radiatif2}
\end{equation}
which allows to consider separately the effects of the source and of the random potential $\mu ({\bf r})$.

\section{Multiple scattering expansion}

Either the expressions (\ref{integralG}) or (\ref{radiatif2}) provide the starting point for a
systematic expansion of the Green function in terms of the free Green function. It can be written
\begin{eqnarray}
G({\bf r},{\bf r}')&=&G_0({\bf r},{\bf r}')+\int d{\bf r}_1
G_0({\bf r},{\bf r}_1) V({\bf r}_1) G_0({\bf r}_1,{\bf r}')
\nonumber \\
&+& \int d{\bf r}_1 d{\bf r}_2 G_0({\bf r},{\bf r}_1) V({\bf r}_1)
G_0({\bf r}_1,{\bf r}_2) V({\bf r}_2) G_0({\bf r}_2,{\bf r}')+
\dots
 \label{devGrr'}
\end{eqnarray}
We can now calculate the average Green function using the white noise potential (\ref{bb}).
 All the odd terms
in the potential $V$ disappear from (\ref{devGrr'}), and it remains
\begin{equation}
\overline{G}({\bf r},{\bf r}') =G_0({\bf r},{\bf r}')+ \gamma \int
d{\bf r}_1 G_0({\bf r},{\bf r}_1) G_0({\bf r}_1,{\bf r}_1)
G_0({\bf r}_1,{\bf r}')+ \dots \label{dev1}
\end{equation}
where we denote from now on the disorder average by
$\bar{.}\bar{.}\bar{.}$~. By averaging over the disorder, the
medium becomes again translational invariant and the Green
function $\overline{G}({\bf r},{\bf r}')=\overline{G}({\bf
r}-{\bf r}')$. The average over the disorder generates all
possible diagrams. Among them, there is a subclass called
irreducible diagrams that cannot be split into two already
existing diagrams without cutting an impurity line. It is
possible to rewrite the average Green function or its Fourier
transform in terms of the contribution of these diagrams only. We
then obtain the so called Dyson equation
\begin{equation}
\overline G({\bf k})=G_0({\bf k})+  G_0({\bf k}) \Sigma({\bf
k},\epsilon) \overline G({\bf k}) \label{dyson}
\end{equation}
where the function $\Sigma({\bf k},\epsilon)$ is called the
self-energy. It should be emphasized that, although the
self-energy contains only the irreducible diagrams, there is an
infinity of them. Thus, the calculation of $\Sigma$ is a
difficult problem. For the white noise potential (\ref{bb}),
$\Sigma$ can be expanded in powers of $\gamma$. To first order,
for the Schr\"odinger equation, we obtain
\begin{equation}
\Sigma_1^{R,A}({\bf k},\epsilon) = {\gamma \over \Omega}
\sum_{\bf q} G^{R,A}_0({\bf q})
\end{equation}
where $\Omega$ is the volume of the system. The real part of $\Sigma$ corresponds to an irrelevant
shift of the origin of the energies that we shall ignore. The imaginary part is
\begin{equation}
\mbox{Im} \Sigma_1^{R}({\bf k},\epsilon) =  -   \pi
\rho_0(\epsilon) \gamma \label{sigma1electron}
\end{equation}
while for the Helmholtz equation, it is
\begin{equation}
\mbox{Im} \Sigma_1^{R}({\bf k},\epsilon) =  - { \gamma  k_0 \over
4 \pi} \label{sigma1ondes}
\end{equation}
We emphasize again that the difference between these two expressions results from
the two distinct dispersions of respectively the Schr\"odinger and Helmholtz equations.
Higher orders terms in the expansion of $\Sigma$ are proportional to $\mbox{Im} \Sigma_1^{R}$ times
some power of the dimensionless parameter $1 \over k l_e$. The contribution $\Sigma_1$ describes
the multiple scattering of the wave as a series of independent effective collisions. The higher
corrections include interference effects between those successive scattering events. The weak disorder
limit $k l_e \gg 1$ amounts to neglecting these interferences. We deal then with the so called self-consistent
Born approximation. We shall, from now on, consider only this limit.

It is then a straightforward calculation to get an expression for the average Green function
at this approximation:
\begin{equation}
\overline{G}^{R,A}({\bf r}_i,{\bf r},k_0)=G_0^{R,A}({\bf r}_i,
{\bf r},k_0) \ \ e^{-|{\bf r} - {\bf r}_i|/2 l_e}
\label{greenmoyen3od2}
\end{equation}
To conclude this section, we would like to notice that although this expression has been obtained
for the case of an infinite system, this restrictive assumption can be released and
we need to consider,
for this relation to be valid, only systems of sizes $L \gg l_e$.

\section{Probability of quantum diffusion}

The quantities of physical interest are usually not related to
the average Green function but instead to the so called {\it
probability of quantum diffusion} which describes the probability
for a quantum particle (or a wave) to go from the point ${\bf r}$
to the point ${\bf r}'$ in a time $t$. Once we average over the
disorder, we shall see that this probability $P( {\bf r}, {\bf
r}', t)$ contains mainly three contributions:
\begin{itemize}
  \item i. The probability to go from ${\bf r}$ to ${\bf r}'$ without
scattering.
  \item ii. The probability to go from ${\bf r}$ to ${\bf r}'$ by an
incoherent sequence of multiple scattering, which is called the
{\it diffuson}.
  \item iii. The probability to go from ${\bf r}$ to ${\bf r}'$ by a
coherent multiple scattering sequence. We shall calculate one
such coherent process, called the {\it cooperon}.
\end{itemize}
We shall first define the probability for the Schr\"odinger case.
Please notice that througout this section we shall take $\hbar =
1$. To that purpose, we consider a gaussian wavepacket of energy
$\epsilon_0$. We shall
 also assume that around $\epsilon_0$
the density of states is constant. Then, we can write for the
Fourier transform $P( {\bf r}, {\bf r}', \omega)$
 of the probability $P( {\bf r}, {\bf r}', t)$ the expression:
\begin{equation}
P({\bf r},{\bf r}',\omega) = {1  \over 2 \pi \rho_0}
 \overline{G^R({\bf r},{\bf r}',\epsilon_0)
 G^A({\bf r}',{\bf r},\epsilon_0-\omega)}
\label{proba1}
\end{equation}
This probability is normalized to unity which means that either
\begin{equation}
\int P({\bf r},{\bf r}',t) d{\bf r}' =1 \label{norma1}
\end{equation}
or
\begin{equation}
\int P({\bf r},{\bf r}',\omega) d{\bf r}' ={i \over \omega}
\label{norma2}
\end{equation}

\subsection{Free propagation}

In the absence of disorder, the Green functions in (\ref{proba1})
 take their free expression (\ref{green3d1}) and it is straightforward
to obtain for the three dimensional case
\begin{equation}
P({\bf r},{\bf r}',t)={\delta(R-v t)\over 4 \pi R^2}
\end{equation}
where $R=|{\bf r}' -{\bf r}|$ and $v$ being the group velocity.
This probability is indeed normalized.

\subsection{Drude-Boltzmann approximation}

In the presence of disorder, we need, in order to calculate the
probability, to evaluate the average of the product of the two
Green functions that appear in (\ref{proba1}).
 The simplest approximation is to replace the average
by the product of the two averaged Green functions. Here again, since we have calculated in the
weak disorder limit (\ref{greenmoyen3od2}) the expression of $\overline G$, we obtain
\begin{equation}
P_0({\bf r},{\bf r}',\omega)={e^{i \omega R /v -R /l_e} \over 4
\pi R^2 v} \label{p0Omega}
\end{equation}
so that
\begin{equation}
\int P_0({\bf r},{\bf r}',\omega) d{\bf r}'= {\tau_e \over 1- i
\omega \tau_e} \label{intP0r2}
\end{equation}
At this approximation, the probability is not normalized, but instead
\begin{equation}
\int P_0({\bf r},{\bf r}',t) d{\bf r}'= e^{- t /\tau_e}
\label{intP0r1}
\end{equation}
It is then clear that some part is missing in the probability. The Drude-Boltzmann approximation
overlooks a large part of the probability; since after a time $t$, it predicts that
 the wavepacket disappears.

\subsection{The diffuson}

There is another contribution associated to the multiple
scattering which can be calculated in the weak disorder limit $k
l_e \gg 1$ in a semiclassical way. Using the description we
obtained previously for the calculation of the average Green
function, we can associate \cite{Chakraverty86} to each possible
sequence $\cal C$, of independent effective collisions a complex
amplitude $A( {\bf r}, {\bf r}', {\cal C})$. Then, using a
generalization of the Feynman path integral description, we can
in principle write the Green function as a sum of such complex
amplitudes.

Then, in order to evaluate the product of two Green functions, we
notice the following two points.
\begin{itemize}
  \item i. Due to the short range of the scattering potential, the set of
scatterers entering in the sequences for both $G^R$ and $G^A$
must be identical.
  \item ii. For the effective collisions, the mean distance between them
is set by the elastic mean free path $l_e \gg \lambda$.
Therefore, if any two scattering sequences differ by even one
collision event, the phase difference between the two complex
amplitudes, which measures the difference of path lengths in units
of $\lambda$ will be very large and then the corresponding
probability will vanish on average.
\end{itemize}
We shall therefore retain only contributions for which the
corresponding probability $P_d({\bf r},{\bf r}',\omega)$ is
\begin{eqnarray}
P_d({\bf r},{\bf r}',\omega)&=&{1 \over 2 \pi \rho_0}\int
\overline G^R_\epsilon ({\bf r},{\bf r}_1) \overline
G^A_{\epsilon-\omega}({\bf r}_1,{\bf r}) \overline G^R_\epsilon
({\bf r}_2,{\bf r}')
\overline G^A_{\epsilon-\omega}({\bf r}',{\bf r}_2) \times \nonumber \\
&\times&\Gamma_\omega({\bf r}_1,{\bf r}_2)d{\bf r}_1 d{\bf r}_2
\label{Pd1}
\end{eqnarray}
It is made of two multiplicative contributions. The first one is
$$\overline G^R_\epsilon ({\bf r},{\bf r}_1)
\overline G^R_\epsilon ({\bf r}_2,{\bf r}') \overline
G^A_{\epsilon-\omega}({\bf r}_1,{\bf r}) \overline
G^A_{\epsilon-\omega}({\bf r}',{\bf r}_2) \ \ .$$ It describes
the mean propagation between
 whatever two points ${\bf r}$ and ${\bf r}'$
in the medium and the first (${\bf r}_1$) (respectively the last
(${\bf r}_2$)) collision event of the
 sequence of scattering events.
The second contribution defines the quantity $\Gamma_\omega({\bf
r}_1,{\bf r}_2)$ which we shall call the {\it structure
factor} of the scattering medium. In a
sense, it generalizes to the multiple scattering
 situation the
usual two-point correlation function in the single scattering
case. We now use once again the assumption of independent
collisions in order to write for $\Gamma_\omega({\bf r}_1,{\bf
r}_2)$ the integral equation
\begin{equation}
\Gamma_\omega({\bf r}_1,{\bf r}_2)=\gamma \delta ({\bf r}_1 -
{\bf r}_2) + \gamma \int \overline G^R_\epsilon ({\bf r}_1,{\bf
r}) \overline G^A_{\epsilon-\omega}({\bf r},{\bf r}_1)
\Gamma_\omega({\bf r},{\bf r}_2) d{\bf r} \label{integgamma}
\end{equation}

This equation can be solved exactly in some geometries. For the infinite three
dimensional space, we can make use
of the translational invariance and get for the structure factor the expression
\begin{equation}
\Gamma_\omega({\bf q})= {\gamma \over 1 - P_0({\bf
q},\omega)/\tau_e} \label{gamcomplet}
\end{equation}
where $P_0({\bf q},\omega)$ is the Fourier transform of
(\ref{p0Omega}) and is given by ${1 \over q v}\arctan {q l_e
\over 1 - i \omega \tau_e}$ with $q=|{\bf q}|$. Then, the
probability rewrites
\begin{equation}
P_d({\bf q},\omega)=P_0({\bf q},\omega) {P_0({\bf
q},\omega)/\tau_e \over 1 - P_0({\bf q},\omega)/\tau_e}
\end{equation}
Using this expression of $P_d$, the normalization of the total
probability $P = P_0 + P_d$ can be readily checked namely $P({\bf
q}=0,\omega)={i\over \omega }$. For the semi-infinite space with
a point source, it is also possible to obtain a closed analytical
expression for the probability $P_d$ using the Wiener-Hopf
method. But beyond these two cases, for simple finite geometries,
it is not possible to obtain the solution of (\ref{Pd1}) without
resorting to numerical calculations. We are then led to look for
some approximate solutions. An excellent one is the
diffusion approximation
obtained for large times $t \gg \tau_e$ and large spatial
variations $r \gg l_e$. It is obtained by expanding the structure
factor under the form
\begin{equation}
\Gamma_\omega({\bf r},{\bf r}_2)=\Gamma_\omega({\bf r}_1,{\bf
r}_2)+({\bf r} - {\bf r}_1).\nabla_{{\bf r}_1} \Gamma_\omega+ {1
\over 2} [({\bf r} - {\bf r}_1).\nabla_{{\bf r}_1}]^2
 \Gamma_\omega
\label{expansion}
\end{equation}
which together with the integral equation (\ref{integgamma}) gives
\begin{equation} \left[ - i \omega  - D\Delta_{{\bf r}_1}   \right]
\Gamma_\omega({\bf r}_1,{\bf r}_2) ={1 \over 2 \pi \rho_0 \tau_e
^2} \delta({\bf r}_1-{\bf r}_2) \end{equation} where the diffusion
coefficient is $D = {1 \over d} {
l_e^2  \over \tau_e} = {1 \over d} v^2 \tau_e$. At this
approximation, we have between $P_d$ and $\Gamma_\omega$ the
following relation \begin{equation} P_d({\bf r},{\bf
r}',\omega)\simeq 2 \pi \rho_0 \tau_e^2 \Gamma_\omega({\bf
r},{\bf r}') \label{Pd9} \end{equation} so that $P_d$, as well,
obeys a diffusion equation.
 It is interesting to check the validity of the diffusion
approximation. For an infinite system, and for ${r \over l_e} =1$ the relative correction between
the exact solution
and diffusion approximation is 0.085 while for ${r \over l_e} = 2.5$, it is less than $5. 10^{-3}$.

\subsection{The cooperon}

With the normalized expression of the probability we just have
obtained, it seems that we fulfilled the demand of evaluating all
the relevant processes that contribute to the probability. But it
could be, and it is certainly the case, that there are many other
contributions that sum up to zero.

For instance, we may consider the possibility which corresponds to the product of Green
functions such as we considered before, but where now the two
identical trajectories are time reversed one from the other. It
is clear that if these trajectories are closed on themselves,
there is no phase difference left between them. This requires
that the system has time-reversal invariance namely that
$G^{R,A}({\bf r},{\bf r}',t) = G^{R,A}({\bf r}',{\bf r},t)$. This
relation does not hold anymore in the presence of a magnetic
field for electronic systems.

The contribution, we shall call $P_c$, of this process to the total probability can be evaluated
as we did before for the diffuson. Thus we have instead of (\ref{Pd1})
\begin{eqnarray}
P_c({\bf r},{\bf r}',\omega)&=&{1 \over 2 \pi \rho_0}\int
\overline G^R_\epsilon({\bf r},{\bf r}_1) \overline
G^R_\epsilon({\bf r}_2,{\bf r}') \overline
G^A_{\epsilon-\omega}({\bf r}',{\bf r}_1)
\overline G^A_{\epsilon-\omega}({\bf r}_2,{\bf r})\times \nonumber \\
&\times&\Gamma'_\omega({\bf r}_1,{\bf r}_2)d{\bf r}_1 d{\bf r}_2
\label{cooperon}
\end{eqnarray}
where the new structure factor $\Gamma'_\omega$ is solution of the integral equation
\begin{equation}
\Gamma'_\omega({\bf r}_1,{\bf r}_2)=\gamma \delta({\bf r}_1-{\bf
r}_2)+ \gamma \int \overline G^R_\epsilon({\bf r}_1,{\bf r}'')
\overline G^A_{\epsilon-\omega}({\bf r}_1,{\bf r}'')
\Gamma'_\omega({\bf r}'',{\bf r}_2) d{\bf r}'' \label{Gammap}
\end{equation}
Notice that unlike the structure factor of the
diffuson, the new combination $\overline G^R_\epsilon({\bf
r}_1,{\bf r}'') \overline G^A_{\epsilon-\omega}({\bf r}_1,{\bf
r}'')$ cannot be simply written in terms of the probability $P_0$.
But as before, we can evaluate $P_c({\bf r},{\bf r}',\omega)$ in
the diffusion approximation (i.e. for slow variations) and we
obtain \begin{equation} P_c({\bf r},{\bf
r}',\omega)\simeq{\Gamma_\omega({\bf r},{\bf r})\over 2 \pi
\rho_0}[\int \overline G^R_\epsilon({\bf r},{\bf r}_1) \overline
G^A_{\epsilon}({\bf r}',{\bf r}_1)d{\bf r}_1 ]^2
\label{cooperon2} \end{equation} Since in the presence of time
reversal invariance, we have
\begin{equation}
\overline G^R_\epsilon({\bf r}_1,{\bf r}) \overline
G^A_\epsilon({\bf r}_1,{\bf r}) = \overline G^R_\epsilon({\bf
r}_1,{\bf r}) \overline G^A_\epsilon({\bf r},{\bf r}_1)
\end{equation}
then, $ \Gamma'_\omega({\bf r}_1,{\bf r}_2) = \Gamma_\omega({\bf
r}_1,{\bf r}_2) $ and finally,
\begin{equation}
P_c({\bf r},{\bf r}',\omega)=P_d({\bf r},{\bf r},\omega)g({\bf
r}-{\bf r}')^2 \label{cooperon22}
\end{equation}
where in $3d$ we have the relation
\begin{equation}
g(R)={\sin k R \over k R} e^{- R /2 l_e}
\end{equation}
and $R = |{\bf r} -{\bf r}'|$. For $R=0$, i.e. for ${\bf r} =
{\bf r}'$, we have
\begin{equation}
P_c({\bf r},{\bf r},\omega)=P_d({\bf r},{\bf r},\omega)
\end{equation}
namely,
the probability to come back to the initial point is {\it twice } the value given by the diffuson.
The contribution of the cooperon $P_c$ to the total probability is given by
\begin{equation}
\int P_c({\bf r},{\bf r}',\omega)  d{\bf r}' = P_d({\bf r},{\bf
r},\omega) { \tau_e \over \pi  \rho_0} \label{pcpd}
\end{equation}
for any space dimensionality. How does this contribution compare with the
diffuson contribution ? We have found that $ P_d ( q= 0, t) \simeq 1$ while
$P_c ( q=0, t) \simeq { \tau_e \over \pi  \rho_0} {1 \over (D t)^{3/2}}$ for small enough times.
Then, $P_c ( q=0, t)$ is maximum for $t \simeq \tau_e$ and given by
$P_c ( q=0, \tau_e) = {1 \over (k l_e )^{d-1}}$. Thus, the contribution of $P_c$ to the total
probability is vanishingly small for $ k l_e \gg 1 $ and for a space dimensionality $d \geq 2$.
But although it is very small, $P_c$ must be compensated by another contribution
in order to restore the normalization of the probability. The additional contributions result
from other irreducible diagrams. But the subsequent terms in this series are not known.

We would like to conclude this section on the cooperon by
emphasizing that although $\Gamma'_\omega$ obeys a diffusion
equation, it would be meaningless and incorrect to state that
$P_c({\bf r},{\bf r}',\omega)$ obeys it as well. The exact
statement is that for $ {\bf r} = {\bf r}'$, $P_c$ and $P_d$ are
proportional and that $P_d$ obeys a diffusion equation.

\section{Radiative transfer}

\subsection{Local intensity and correlation function}

In the previous sections, we have defined the quantum probability for electronic systems.
The probability $P$
 is directly related to quantities that are physically measurable like the electrical
conductivity or the magnetic response e.g. the magnetization \cite{gilles1,gilles2}.
 For the study of the
propagation of electromagnetic waves in disordered media, the quantity which is usually measured
is the local intensity of the field or its correlation function \cite{ishimaru}. As we discussed previously,
 and by definition of the Green function, the radiative solution
$\psi_{\epsilon} ({\bf r})$ of the Helmholtz equation
(\ref{radiatif})
 with a localized source at point ${\bf R} = {\bf 0}$ is
${\overline \psi}_{\epsilon} ({\bf r}) =  G_{\epsilon}({\bf
0},{\bf r}) $. The correlation function of the field is then
\begin{equation}
\overline {\psi_{\epsilon} ({\bf r}) \psi^*_{ \epsilon - \omega}
({\bf r}')} = \overline {G^R_{\epsilon} ({\bf 0},{\bf r})
G^A_{\epsilon - \omega} ({\bf r}', {\bf 0})} \label{correlation1}
\end{equation}
It is not directly related to the probability of quantum
diffusion $P ({\bf r}, {\bf r}', \omega)$. But the radiated
intensity $I ({\bf r})$ defined by
\begin{eqnarray}
I({\bf r}) & = & { 4 \pi \over c} |\psi_{\epsilon}({\bf r}){|^2} \nonumber \\
& = &{4 \pi \over c}  {G^R_{\epsilon} ({\bf 0},{\bf r})
G^A_{\epsilon} ({\bf r}, {\bf 0})} \label{intensite}
\end{eqnarray}
is indeed related on average to the probability $P$ and
\begin{equation}
\overline{I}({\bf r}) ={ 4 \pi \over c} \overline{G^R_{\epsilon}
({\bf 0},{\bf r}) G^A_{\epsilon} ({\bf r}, {\bf 0})}
\end{equation}
Using for the probability the following relation which is the counterpart, for
 the Helmholtz equation, of the relation (\ref{proba1})
\begin{equation}
P_d({\bf r},{\bf r}') ={ 4 \pi \over c} \overline{G^R_{\epsilon}
({\bf r},{\bf r}') G^A_{\epsilon} ({\bf r}', {\bf r})}
\end{equation}
we obtain $\overline{I}({\bf r}) = P_d({\bf 0},{\bf r})$. From
now on, we shall denote $I ({\bf r})$ the average intensity.

We can rephrase what we did before in order to calculate the various contributions to the
intensity that come respectively from the Drude-Boltzman, the diffuson and the cooperon approximations.
The first contribution is given by
\begin{equation}
I_0(R)= {1 \over 4 \pi R^2 c}e^{-R /l_e}
\end{equation}
It corresponds to the contribution to the radiative intensity of waves that did not experience
any collision on a distance $R$ from the source.

The diffuson contribution is given by
\begin{equation}
I_{d}({\bf r}) = {4 \pi \over c} \int d {\bf r}_1 d {\bf r}_2
 |{\overline
\psi}_{\epsilon}({\bf r}_1)|^2 \Gamma_{\omega=0}({\bf r}_1, {\bf
r}_2)|{\overline G}_{\epsilon}^R ({\bf r}_2 , {\bf r}) |^2
\label{intensited}
\end{equation}
and finally, the contribution of the cooperon to the intensity is
\begin{equation}
I_{c}({\bf r}) = {4 \pi \over c} \int d {\bf r}_1 d {\bf r}_2
{\overline \psi}_{\epsilon}({\bf r}_1) {\overline
\psi}^*_{\epsilon}({\bf r}_2) \Gamma({\bf r}_1, {\bf r}_2)
{\overline G}_{\epsilon}^R ({\bf r}_2 ,{\bf r}) {\overline
G}_{\epsilon}^A ({\bf r} , {\bf r}_1) \label{intensitec}
\end{equation}
where we used the notation $\Gamma_{\omega=0} = \Gamma$.
In the diffuson approximation, the intensity $I_d$ rewrites
\begin{equation}
I_d({\bf r}) =P_d({\bf 0},{\bf r})= {l_e^2 \over 4 \pi c}
\Gamma({\bf 0},{\bf r})
\end{equation}
and like $P_d$, it obeys the diffuson equation
\begin{equation}
 - D\Delta I_d({\bf r}) = \delta({\bf r})
\end{equation}
whose solution in the $3d$ free space is
\begin{equation}
I_d(R)={1 \over 4 \pi D R}
\end{equation}

We shall now apply all the considerations developed in this
section to the calculations of physical quantities in some
specific situations both for metallic systems and for the
propagation of electromagnetic waves in suspensions.

\section{Example 1. The electrical conductivity of a weakly disordered metal}

We previously defined a weakly disordered metal as a non
interacting and degenerate (spinless)
 electron gas
(at $T = 0$), moving in the field of defects  and impurities described by the white
 noise potential (\ref{bb}).

The average electrical conductivity ${\sigma}(\omega)$ calculated in the framework of the linear response theory
 \cite{Doniach74} is given
by the Kubo formula:
\begin{equation}
{\sigma}(\omega) = {{{e^2} {\hbar^3} } \over { 2 \pi {m^2}\Omega}}
\sum_{{\bf k},{\bf k}'} {k_{x}} {k'_{x}} \ \ \overline
{{G_{\epsilon}^R} ({\bf k},{\bf k'}) {G_{\epsilon - \omega}^A}
({\bf k'},{\bf k}) } \label{kubo}
\end{equation}

Then, we see from this definition, that the structure of the conductivity
is up to the product ${k_{x}} {k'_{x}}$ very similar to those of the quantum probability $P$.
 Therefore, and just as we did before, the very definition of ${\sigma}(\omega)$
 leads us to the following set of approximations.

\subsection{ The Drude-Boltzman approximation}

It is obtained by replacing the average of the product of two Green functions by the product
of the averages, namely
\begin{equation}
\overline {{G_\epsilon^R} ({\bf k},{\bf k'})
 {G_{\epsilon- \omega}^A} ({\bf k'},{\bf k}) } \simeq
 {{{\overline G}_\epsilon^R} ({\bf k}, {\bf k}')
 {{\overline G}_{\epsilon- \omega}^A ( {\bf k}', {\bf k}) }}
\end{equation}
where the Fourier transform of the averaged Green functions (\ref{greenmoyen3od2}) is
\begin{equation}
{\overline G}_\epsilon^{R,A} ({\bf {k}}, {\bf {k'}}) = {\overline
G}_\epsilon^{R,A} ({\bf k}) \delta_{{\bf k},{\bf k}'} = {
\delta_{{\bf k},{\bf k}'} \over { \epsilon - \epsilon({\bf k})
\pm i {\hbar \over {2 \tau_e}}}}
\end{equation}
Then, we obtain for the conductivity $\sigma_0 (\omega)$ at this approximation the following
expression
\begin{equation}
 \sigma_0 (\omega) = {{n e^2 } \over m}
\int d {\bf r}' P_0 ({\bf r}, {\bf r}', \omega)
\end{equation}
where $P_0$ is the quantum probability calculated at the same approximation (\ref{p0Omega}).
 Using the expression
(\ref{intP0r2}), we obtain
\begin{equation}
\sigma_0 (\omega) = {{n e^2 } \over m} { \tau_e \over {1 - i \omega \tau_e}}
\end{equation}
which is the well-known Drude expression. It must be noticed that
because of the Kronecker delta function that appears in the
average Green functions, the scalar product ${k_{x}} {k'_{x}}$
reduces simply to $k_{x}^2$ and eventually, after averaging, to
$k_F^2  \over d$ where $k_F$
 is the Fermi wavevector.

\subsection{The contributions of the diffuson and the cooperon}

Here again, we approximate the average product in the relation (\ref{kubo}), using the same
scheme we used for the diffuson. The contribution $\sigma_d (\omega)$ of the diffuson to the
conductivity is thus
\begin{equation}
 \sigma_d (\omega) = {{{e^2} {\hbar^3} } \over { 2 \pi {m^2} \Omega^2}}
{1 \over 2 \pi \rho_0 \tau_e^2} P_d ({\bf 0}, \omega) \sum_{{\bf
k},{\bf k}'} {k_{x}} {k'_{x}}  \tilde P_0 ({\bf k} ,{\bf
0},\omega) \tilde P_0 ({\bf k}',{\bf 0} ,\omega)
\end{equation}
where the function $\tilde P_0 ({\bf k}, {\bf q} ,\omega) = {1
\over 2 \pi \rho_0} {{\overline G}_{\epsilon}^R} ({\bf k} + {{\bf
q} \over 2}) {{\overline G}_{\epsilon - \omega}^A} ({\bf k} -
{{\bf q} \over 2})$. Since the function $\tilde P_0 ({\bf k},
{\bf q} = {\bf 0},\omega)$ depends only on the modulus of the
wavevector ${\bf k}$ and not on his direction, the angular
integral in the previous expression gives a vanishing
contribution namely $\sigma_d (\omega) = 0$. Then, it is
interesting to notice that although the diffuson gives the main
contribution to the quantum probability, its contribution to the
conductivity which indeed measures such a probability, vanishes
identically.

We evaluate, the contribution $\sigma_c (\omega)$ of the cooperon using the relation (\ref{cooperon})
so that
\begin{equation}
  \sigma_c (\omega) = - {{{e^2} {\hbar^3} } \over { 2 \pi {m^2}}} {{k_F^2} \over d}
 \big[ {1 \over \Omega} \sum_{{\bf k}}  \tilde P_0^2 (k,{\bf q} = {\bf 0}, \omega) \big]
 {1 \over \Omega} \sum_{\bf Q} P_d ({\bf Q}, \omega)
\end{equation}
the sum in the brackets is straightforward so that in the diffusion approximation, we obtain
\begin{equation}
 \sigma_c (\omega) = -{{e^2} D  \over {  \pi \hbar}} P_d({\bf r},{\bf r},\omega)
\label{deltasigma11}
\end{equation}
Using now the relation (\ref{pcpd}) between the cooperon and the diffuson, $\sigma_c (\omega)$
rewrites
\begin{equation}
 \sigma_c (\omega) = - {n e^2 \over m} \int d{\bf r}' P_c ({\bf r}, {\bf r}', \omega)
\end{equation}
and the total conductivity at this order is now given by
\begin{equation}
\sigma (\omega) = {n e^2 \over m} \int d{\bf r}' \big( P_0 ({\bf
r}, {\bf r}', \omega) - P_c ({\bf r}, {\bf r}', \omega) \big)
\label{sigma1}
\end{equation}

The conductivity $\sigma (\omega)$ is reduced by the coherent (cooperon) contribution.
This correction is called the {\it weak localization correction} to the conductivity
\cite{Bergmann84,Rammer99,houches94}.
 Relatively,
this contribution of $P_c$ is much larger than the normalization correction to the total
probability. This is because $P_c$ is now compared to $P_0$ and not to $P_d$
 which represents the main
contribution.

\subsection{The recurrence time}

It is of some interest to study the dc conductivity $\sigma
(\omega =0)$ using a slightly different point of view \cite{am}.
>From the relation (\ref{sigma1}), and using the dc expression
$\sigma_0 = {n e^2 \tau_e \over m}$, we obtain for the relative
correction to the conductivity,
\begin{equation}
{\delta \sigma \over \sigma_0} = - {1 \over \pi \hbar \rho_0 }
\int_{0}^{\infty} dt P_c ({\bf r}, {\bf r}, t) \label{relative}
\end{equation}
where $\rho_0$ is the density of states per unit volume.

Consider now the quantity $Z(t)$ defined by
\begin{equation}
Z(t) = \int_\Omega P_d({\bf r},{\bf r},t) d{\bf r}
\end{equation}
where the integral is over the volume $\Omega$ of the system. It
represents the return probability to a point ${\bf r}$ averaged
over all those points.
 This quantity which characterizes the solutions of the diffusion  equation
is sometimes called {\it the heat kernel} in the literature. The time integral of $Z(t)$
 defines the characteristic time $T_R$
\begin{equation}
T_R  =  \int_\Omega d{\bf r} \int_0^\infty  P_d ({\bf r},{\bf
r},t) dt  = \int_0^\infty Z(t) \ dt \label{TRglobal}
\end{equation}
called the {\it recurrence time}. It measures the space average of the time spent by a diffusive particle
within each infinitesimal volume. $T_R$ diverges, as stated by the Polya theorem \cite{polya}, for a random walk
in the free space of dimensionality $d \leq 2$. Then, we define a
regularized expression for $T_R$ given by the Laplace transform
\begin{equation}
T_R(s)  = \int_0^\infty Z(t)  e^{- s t}\ dt
\label{tgamma}
\end{equation}
This expression of the recurrence time may be interpreted by
saying that $T_R(s)$ selects the contribution of all the
diffusive trajectories of lengths smaller than $L_s = \sqrt {D/
s}$.

Then, the expression (\ref{relative}) of the relative contribution of the cooperon to the
conductivity rewrites
\begin{equation}
{\delta \sigma \over \sigma_0} = - {\Delta \over \pi \hbar}  \int_{0}^{\infty} dt Z(t) e^{- s t}
=  - {\Delta \over \pi \hbar} T_R(s)
\label{sigmaz}
\end{equation}
where the energy defined by $\Delta = {1 \over \Omega \rho_0}$ is the mean level spacing measured
at the Fermi level between the energy levels of an electron gas confined in a box of volume $\Omega$.
The weak localization correction to the dc conductivity can be essentially expressed in terms of
a purely classical quantity, namely the recurrence time $T_R$. When it diverges, we need to use its
regularized version (\ref{tgamma}), where now the length $L_s$ can be given a physical meaning.
 The cooperon correction results from the absence of any relative phase between two time reversed
multiple scattering trajectories. Any interaction of an electron with an external perturbation may
destroy this phase coherence and then the contribution of the cooperon. For a large class of
such perturbations which includes inelastic collisions at finite temperature, Coulomb interactions or
coupling to other excitations, we can define a phenomenological length $L_\phi$ called the phase
coherence length and a phase coherence time $\tau_\phi$ such that $L_\phi^2 = D \tau_\phi$.
Thus, $L_\phi$ is the length $L_s$  used to regularize the recurrence time.
We shall see later other examples of physical quantities that can be expressed using either the
heat kernel or the recurrence time.

\subsection{Conductance fluctuations- Spectral determinant}

We could as well calculate the so-called conductance fluctuations
using the properties of the heat kernel. The conductance $G$ is
related to the conductivity $\sigma$ by the Ohm's law $G = \sigma
L^{d-2}$. We define the dimensionless conductance $g = { h \over
e^2} G$. Its fluctuation defined by $\langle \delta g^2 \rangle =
\langle g^2 \rangle - \langle  g \rangle^2$ can be expressed in
terms of $Z(t)$ through \begin{equation} \langle \delta g^2
\rangle = {12 \over \tau_D^2} \int_{0}^{\infty} dt t Z(t) e^{- s
t} \end{equation} where $\tau_D = L^2/D$ is the diffusion time.
For a one dimensional system (in the sense of the diffusion
equation), we have \begin{equation} \langle \delta g^2 \rangle =
{2 \over 15} \end{equation} It is interesting to recover the last
two results on the average conductivity (or conductance) and its
fluctuation using a systematic expansion of the recurrence time.
To that purpose, we define the spectral determinant $S(s)$ by
$S(s) = det( - \Delta + s) $ where $\Delta$ is the Laplacian
operator. Then, by definition of $Z(t)$, we have the relation
\begin{equation} T_R(s) = \int_0^\infty Z(t)  e^{- s t}\ dt =
{\partial \over
\partial s} \mbox{ln} S(s) \end{equation} up to a regularization
independent of the Laplace variable $s$. This relation is valid
for all space dimensionality. Consider now as a working example
\cite{gilles2}
 the case of a $1d$ diffusive wire of
length $L$. In order to describe the case of a perfect coupling
of the wire to the reservoir, we demand Dirichlet boundary
conditions for the diffusion equation. This is in contrast to the
Schr\"odinger equation for which this choice corresponds to
Neumann boundary conditions. Then, the spectral determinant
$S(s)$ of the diffusion equation can be readily calculated and it
is given by \begin{equation} S(x) = {\sqrt x \over \sinh \sqrt x}
\end{equation} with $x = s \tau_D$. The average conductance
$\delta g$ deduced from (\ref{sigmaz}) and the fluctuation
$\langle \delta g^2 \rangle$ can be written generally as
\begin{eqnarray}
\delta g & = &  2 {\partial \over \partial x} \mbox{ln} S(x) = - {1 \over 3} \\
\langle \delta g^2 \rangle & = &  12
{\partial^2 \over \partial x^2} \mbox{ln} S(x) = {2 \over 15}
\end{eqnarray}
In the limit $x \rightarrow 0$, we can expand the spectral determinant namely
$\mbox{ln} S(x) \simeq_{x \rightarrow 0}  - {x \over 6} + {x^2 \over 180}$
and recover the previous result.

\section{Example 2. Multiple scattering of light: the albedo}

We shall now give an example of the use of the quantum probability
taken from the multiple scattering of electromagnetic waves in disordered
suspensions. To that purpose, consider the scattering medium as being the half-space $z \geq 0$.
The other half-space is a free medium which contains both the sources of the waves and the detectors.
We also assume that both incident and emergent waves are plane waves with respective wavevectors
${\bf k_i} = k {\bf \hat s_i}$ and $ {\bf k_e} = k {\bf \hat s_e}$, where ${\bf \hat s_i}$  and
${\bf \hat s_e}$ are unit vectors. The waves experience only elastic scattering in the medium
 so that after a
collision, only the direction $\bf \hat s$ changes while the
amplitude $k = {\omega_0 \over c}$ remains constant. The
reflection coefficient $\alpha ({\bf \hat s_i}, {\bf \hat s_e})$
for this geometry is called the {\it albedo} and is proportional
to the intensity $I({\bf R}, {\bf \hat s_i}, {\bf \hat s_e})$
emerging from the medium per unit surface and per unit solid
angle measured at a point $\bf R$ at infinity. If $F_{inc}$
defines the flux of the incident beam (related to the incident intensity $I_{inc}$ 
by $F_{inc} = c S I_{inc}$) where $S$ is the illuminated
surface in the plane $z =0$, then the albedo is given by
\begin{equation}
\alpha ({\bf \hat s_i}, {\bf \hat s_e})  =
{R^2 c \over F_{inc} } I({\bf R},{\bf \hat s_i}, {\bf \hat s_e})
\label{albedo}
\end{equation}
In order to calculate the intensity $I({\bf R}, {\bf \hat s_i}, {\bf \hat s_e})$, we need to calculate
first the radiative solutions of the Helmholtz equation at the Fraunhoffer approximation. It is
given by
\begin{equation}
\psi_{\omega_0} ({\bf \hat s_i}, {\bf \hat s_e}) = \int d{\bf r}
d{\bf r}' e^{ik ({\bf \hat s_i}.{\bf r} - {\bf \hat s_e}.{\bf
r}')} G({\bf r}, {\bf r}', \omega_0) \label{fraunhoffer1}
\end{equation}
and the intensity is then
$I({\bf R},{\bf \hat s_i}, {\bf \hat s_e}) = |\psi_{\omega_0} ({\bf \hat s_i}, {\bf \hat s_e})|^2$. Before
averaging over the disorder, the albedo looks like a random pattern of bright and dark spots.
This is a speckle pattern \cite{dainty,maret94} whose long range correlations
 are a specific feature of multiple scattering. By averaging over the disorder
(either liquid suspension \cite{Kuga84,Vanalbada85,Wolf85} or rotating solid sample \cite{Kaveh86}),
 the speckle is washed out and the two
surviving contributions are respectively associated to the
diffuson and the cooperon.

\subsection{The incoherent albedo: the diffuson}

We have calculated previously the various contributions to the
intensity. The incoherent part is given by the diffuson contribution, namely the relation
(\ref{intensited}), where we take as the source term the incident plane wave
\begin{equation}
{\overline \psi_i}({\bf r}_1) = \sqrt{{ c I_{inc} \over 4 \pi}}
e^{- {|{\bf r}_1 - {\bf r}| /2 l_e}}
 e^{- i k {\bf \hat s_i}. {\bf r}_1}
\label{source}
\end{equation}
where ${\bf r}$ is the impact point on the surface $z =0$ and
${\bf r}_1$ is the position of the first collision. The albedo
rewrites
\begin{equation}
\alpha_d = {R^2 \over  S} \int d{\bf r}_1 d{\bf r}_2 |{\overline
G}^R ({\bf r}_2 , {\bf R}) |^2 \Gamma ({\bf r}_1, {\bf r}_2) e^{-
{|{\bf r}_1 - {\bf r}| / l_e}} \label{albedod}
\end{equation}
The Fraunhoffer approximation consists in taking the limit $|{\bf
R} - {\bf r}_2| \rightarrow \infty$, so that the Green function
${\overline G^R}({\bf r}_2, {\bf R})$ can be expanded as
\begin{eqnarray}
{\overline G^R}({\bf r}_2, {\bf R}) & = & e^{-|{\bf r}'-{\bf
r}_2|/ 2 l_e}
{e^{ik |{\bf R} - {\bf r}_2|} \over 4 \pi |{\bf R} - {\bf r}_2|} \nonumber \\
& \simeq &  e^{-|{\bf r}'-{\bf r}_2|/ 2 l_e}
 e^{- i k {\bf \hat s_e}. {\bf r}_2 } {e^{-ik  R} \over 4 \pi  R}
\label{greenfraun}
\end{eqnarray}
By defining the respective projections $\mu$ and $\mu_0$ of the vectors ${\bf \hat s_i}$ and
${\bf \hat s_e}$ on the $z$ axis, we obtain finally
\begin{equation}
\alpha_d = {1 \over (4 \pi )^2 S} \int d{\bf r}_1 d{\bf r}_2 e^{-
{z_1 \over \mu_0 l_e}}  e^{- {z_2 \over \mu l_e}} \Gamma ({\bf
r}_1, {\bf r}_2)
\end{equation}
This expression calls for a number of remarks. We first notice that it does not contain any dependence
on the direction namely on the incident and emerging vectors ${\bf \hat s_i}$ and ${\bf \hat s_e}$.
Therefore, at this approximation, the albedo is flat i.e. does not have any angular structure.
Second, if we compare this expression to its counterpart for the conductivity, we see that
although both of them describe transport, the conductivity vanishes for the diffuson while it does
not for the albedo.

The structure factor $\Gamma$ is related to the diffusion probability through the relation
\begin{equation}
P_d ({\bf r}_1, {\bf r}_2) = {l_e^2 \over 4 \pi c} \Gamma ({\bf
r}_1, {\bf r}_2)
\end{equation}
so that we obtain
\begin{equation}
\alpha_d = { c \over 4 \pi l_e^2} \int_{0}^{\infty} dz_1 dz_2
e^{- {z_1 \over \mu_0 l_e}}  e^{- {z_2 \over \mu l_e}} \int_{S}
d^2 \rho P_d (z_1, z_2, \rho) \label{albedod1}
\end{equation}
where now due to the geometry of the medium, the function $P_d$
depends on $z_1$, $z_2$ and the projection $\rho$ of the vector
${\bf r}_1 - {\bf r}_2$ onto the plane $z =0$. Within the
diffusion approximation, $P_d$ can be calculated for this
geometry using the image method \cite{Akk86,Akk288}.
 This gives
\begin{equation}
P_d ({\bf r}, {\bf r}') = {1 \over 4 \pi D} \big[ {1 \over |{\bf
r} - {\bf r}'|} -  {1 \over |{\bf r} - {{\bf r}'}^*|} \big]
\label{image}
\end{equation}
where ${{\bf r}'}^* = ({r'}^*_\perp,- z' + 2 z_0)$ and $z_0 = {2
\over 3} l_e$. Then, it appears that $P_d$ does not vanish on the
plane $z = 0$ but instead for $z = - z_0$. It is possible for that
geometry to compare the validity of the diffusion approximation
with the known exact solution for the semi-infinite problem
without sources (the so called Milne problem) that can be solved
using the Wiener-Hopf method \cite{vdh}. There, we obtain instead
that $P_d$ vanishes on the plane $z = - 0.7104 l_e$. This
justifies the use of the diffusion approximation. But, it is
important at this point to make two important remarks. First, it
would be a mistake to believe that on the basis of the two
examples, infinite and semi-infinite spaces, for which the
diffusion approximation works well that it is indeed always the
case. The generalization and the validity of the diffusion
equation in a restricted geometry is a difficult problem. Second,
all the expressions we have obtained so far are valid for
isotropic scatterers for which there is no difference between the
elastic and the transport mean free paths. When relaxing this
approximation, namely dealing with anisotropic scattering,
 we have to include both of them. Although it
appears clearly from simple physical considerations that in the expression (\ref{albedod1}), we must leave
the elastic mean free path $l_e$ in the exponential factors that describe the first and last collisions,
and replace it by the transport mean free path in the expression of $P_d$, this point needs further
investigation using for instance transport theory \cite{ishimaru,Akk288}.

Finally, by replacing (\ref{image}) into (\ref{albedod1}), we obtain
\begin{equation}
\alpha_d = {3 \over 4 \pi} \mu \mu_0 \big( {z_0 \over l_e} + {\mu \mu_0 \over \mu  + \mu_0 }\big)
\end{equation}

\subsection{The coherent albedo: the cooperon}

Along the same lines, we can now evaluate the contribution of the cooperon to the albedo.
Using the expression (\ref{intensitec}) for the intensity,  we obtain
\begin{equation}
\alpha_c ({\bf \hat s_i},{\bf \hat s_e}) =
 {1 \over (4 \pi )^2 S} \int d{\bf r}_1 d{\bf r}_2 e^{- {1 \over 2} ({1 \over \mu} + {1 \over \mu_0})
{z_1 + z_2 \over l_e}}\Gamma ({\bf r}_1, {\bf r}_2) e^{i k ({\bf
\hat s_i } + {\bf \hat s_e} ). ({\bf r}_2 - {\bf r}_1)}
\label{albedoc}
\end{equation}
Unlike the previous case, there is now a phase term present in
this expression which is at the origin of a new angular structure
of the albedo. It is straightforward to check that, assuming
${\bf \hat s_i } + {\bf \hat s_e} = {\bf 0}$, i.e. measuring the
albedo right
 in the backscattering
direction, we have
\begin{equation}
\alpha_c (\theta =0) = \alpha_d
\end{equation}
where the angle $\theta$ is between the directions ${\bf \hat s_i }$ and ${\bf \hat s_e }$.
Then, the total albedo $ \alpha (\theta) = \alpha_d + \alpha_c (\theta)$ is such that
\begin{equation}
\alpha(\theta =0 ) = 2 \alpha_d
\end{equation}
Using the
expression (\ref{image}) for the probability and integrating, we obtain
\begin{equation}
\alpha_c (\theta) = {3  \over 8 \pi } {\mu \mu_0 \over (1 + \mu k_\perp l_e)(1 + \mu_0 k_\perp l_e)}
\big( {1 - e^{- 2 k_\perp z_0} \over k_\perp l_e} + 2 { \mu \mu_0 \over \mu + \mu_0} \big)
\end{equation}
in this expression,
$ {\bf k_\perp} =({\bf k_i}+{\bf k_e})_\perp=k ({\bf \hat s_i} + {\bf \hat s_e})_\perp$ is
the projection onto the plane $xOy$ of the vector ${\bf k_i}+{\bf k_e}$. At small angles, we have
$k_\perp \simeq { 2\pi \over \lambda} \theta$, while at large angles, $\alpha_c
(\theta \rightarrow \infty) =0$. Then, the coherent contribution to the albedo is finite only
in a cone of angular aperture $\lambda \over 2 \pi l_e$ around the backscattering direction.
By expanding near $\theta = 0$ we obtain the expression
\begin{equation}
\alpha_c (\theta)  \simeq \alpha_d - {3 \over 4 \pi}{(l_e + z_0 {)^2} \over l_e}  k_\perp +
 O(k_\perp{)^2}
\end{equation}
Thus, the albedo shows a cusp near $\theta =0$ namely, its derivative is discontinuous at this point.

\section{Example 3: Spectral correlations}

In the previous two examples, we discussed transport
coefficients. They give a description of the system when it is
connected to another ``reference'' medium. There is another
characterization independent of the coupling to another system,
which focuses on {\it spectral properties} i.e. properties of the
energy spectrum of the solutions of the wave equations. For
electronic systems, they are related to the equilibrium
thermodynamic properties like the magnetization of a metallic
sample. For disordered systems, it also raises an important
issue: these systems are part of the larger class of {\it complex
systems}, usually non integrable. The complete set of correlation
functions of the energy levels provides a complete description of
the thermodynamic properties. For instance, it is possible just
by inspection of the energy spectrum to determine whether or not
an electronic system is a good metal or an Anderson insulator.

Let us start by defining the set of eigenenergies $\epsilon_\alpha$ of the Schr\"odinger (or
Helmholtz) equation in a confined geometry (e.g. a box). The density of states per unit volume is
given by
\begin{equation}
\rho(\epsilon) = {1 \over \Omega} \sum_\alpha \delta(\epsilon -\epsilon_\alpha)
\end{equation}
It can be written as well in terms of the Green function for the Schr\"odinger equation, using the
equality
\begin{equation}
\rho(\epsilon) = -{1 \over \pi \Omega} \mbox{Im} \int d{\bf r}
G^R({\bf r},{\bf r},\epsilon) \label{dos}
\end{equation}
By averaging over the disorder, we define the various correlation functions.
For instance, the two-point correlation function is
\begin{equation}
\overline{ \rho(\epsilon_1)\rho(\epsilon_2) }- \overline
\rho(\epsilon_1) \  \overline \rho(\epsilon_2) \label{KE}
\end{equation}
where the average density of states coincides with its free value $\rho_0$ and
therefore is related to the mean level spacing $\Delta$ through
$\overline{\rho}= {1 \over \Delta \Omega}$.
It is important at this point to make the following remark. The average we consider here is over the
random potential (\ref{bb}). We could have taken as well the average over different parts of the
spectrum as is done for instance in the quantum description of
classically chaotic billiards where there is no random potential to average
over. In the diffusion approximation, it has been shown numerically that these two ways to average
are equivalent \cite{gilles1,Bohigas91} but this not need to be the general case. We then define the
dimensionless correlation function
\begin{equation}
K(\omega)= {\overline{  \rho(\epsilon)\rho(\epsilon-\omega) }\over \overline \rho(\epsilon)^2}-1
\label{KE2}
\end{equation}
It can be expressed in terms of the Green function under the form
\begin{equation} K(\omega)={\Delta^2}  \int d{\bf r} d{\bf r}'
K({\bf r},{\bf r}',\omega) \label{KdeE3} \end{equation} with
\begin{equation} K({\bf r},{\bf r}',\omega)={1 \over 2 \pi^2 }
\mbox{Re} \overline{G^R({\bf r},{\bf r},\epsilon) G^A({\bf
r}',{\bf r}',\epsilon-\omega)}^c \label{KdeE4} \end{equation}
where $\overline{...}^c$ represents the cumulant average. The
diffuson and cooperon contributions to the
 local function $K({\bf r},{\bf r}',\omega)$ can be expressed in terms of $P_d$ and the structure factor as
\cite{Althuler86} \begin{equation} K_d({\bf r},{\bf
r}',\omega)={1\over  2 \pi^2 } \mbox{Re} \left[ P_d({\bf r},{\bf
r}'\omega)P_d({\bf r}',{\bf r},\omega)\right] , \end{equation}
while \begin{equation} K_c({\bf r},{\bf r}',\omega)= 2 \rho_0^2
\tau_e^4 \mbox{Re} \left[\Gamma'_\omega({\bf r},{\bf
r}')\Gamma'_\omega({\bf r}',{\bf r})\right] \label{KcKc}
\end{equation} when the system has time-reversal invariance.
Finally, we define the Fourier transform $\tilde{K}(t)$ which is
often called the {\it form factor}. By
 collecting the two previous contributions, it can be written
\begin{equation} \tilde K(t)=   {\Delta^2 \over  4 \pi^2 } |\ t \
| Z( |t|) ={\Delta^2 \over  4 \pi^2 } | t |\  \int_\Omega P_d
({\bf r},{\bf r}, |t|) d{\bf r} \end{equation} which is precisely
the form that is obtained assuming the Random Matrix Theory
 \cite{gilles1,Bohigas91} provided we consider the regime for which
$\hbar / \Delta \gg t \gg \tau_D$ where
$\tau_D = L^2 / D$. The first inequality which involves the Heisenberg time $\hbar / \Delta$
enforces the condition of a continuous spectrum while the second one indicates that we
must be in the ergodic limit where the diffusing particle explored the whole system.

\section{Conclusion}

We have presented in this short review a very partial selection
of highlights in the field of multiple scattering of waves in
disordered media. Our aim was more to give a feeling of the
profound unity of the physical phenomena and therefore of the
methods to handle them rather than
 discussing an extensive list of effects. Among them, we should mention the study of fluctuations
of the intensity for waves both in the weak disorder regime and near the Anderson transition that
will be covered in details by the reviews of R. Pnini, A. Genack and G.Maret.
This is certainly a problem
for which the recent developments have been very spectacular. We have studied the coherent albedo in
the weak disorder case. It should  be extended to the strong disorder limit.
Finally, our last example on spectral correlations should be considered as
 another way to obtain the semiclassical results reviewed in the contribution
of D. Delande.


\begin{thebibliography}{article}

\bibitem{am} These notes are a partial and preliminary account of the monography
{\it Coh\'erence et diffusion dans les milieux d\'esordonn\'es} by E. Akkermans and
G. Montambaux, to be published by CNRS Intereditions for the french version.

\bibitem{chandra} S. Chandrasekhar, {\it Radiative transfer} (Dover, N.Y. 1960).

\bibitem{Chakraverty86}S. Chakraverty and A. Schmid,
Phys. Rep. {\bf 140},193 (1986)

\bibitem{gilles1} G. Montambaux, in {\it Quantum fluctuations}, proceedings of the
Les Houches Summer School, Session LXIII, ed. by S. Reynaud et al.
(North Holland, Amsterdam, 1996), p.387.

\bibitem{gilles2} M. Pascaud and G. Montambaux, Phys. Rev. Lett. $\bf 82$ (1999), 4512
and Phys. Uspekhi {\bf 41},182 (1998).

\bibitem{ishimaru} A. Ishimaru, {\it Wave propagation and scattering in random media},
Vol.I,II (Academic, N.Y. 1978).

\bibitem{Doniach74} S. Doniach et E.H. Sondheimer,
{\it Green's Functions for Solid State Physicists},
Frontiers in Physics, W.A. Benjamin, (1974)

\bibitem{Bergmann84}G. Bergmann, Phys. Rep. {\bf 107}, 1 (1984)

\bibitem{Rammer99}J. Rammer, {\em Quantum transport theory}, Frontiers in Physics,
 Perseus books (1998)

\bibitem{houches94} {\it Mesoscopic quantum physics}, proceedings of the
Les Houches Summer School, Session LXI, ed. by E.Akkermans, G. Montambaux, J.L. Pichard and
J. Zinn-Justin (North Holland, Amsterdam, 1995).

\bibitem{polya} C. Itzykson and J.M. Drouffe, {\it Statistical field theory}, Vol. 1 (Cambridge, 1989)

\bibitem{dainty} C. Dainty ed., {\it Laser speckle and related phenomena}, Topics in a
applied physics, Vol.9 (Springer, Berlin, 1984).

\bibitem{maret94} G. Maret, in {\it Mesoscopic quantum physics}, proceedings of the
Les Houches Summer School, Session LXI, ed. by E.Akkermans, G. Montambaux, J.L. Pichard and
J. Zinn-Justin (North Holland, Amsterdam, 1995).

\bibitem{Kuga84} Y. Kuga and A. Ishimaru, J.Opt.Soc. Am., {\bf A8}, 831 (1984)

\bibitem{Vanalbada85} M.P. van Albada et A. Lagendijk, Phys. Rev. Lett. $\bf 55$ (1985), 2692.

\bibitem{Wolf85} P.E. Wolf et G. Maret, Phys. Rev. Lett. $\bf 55$ (1985), 2696.

\bibitem{Kaveh86} M. Kaveh, M. Rosenbluh, I. Edrei et I. Freund, Phys. Rev. Lett. $\bf 57$ (1986), 2049

\bibitem{Akk86} E. Akkermans, P.E. Wolf, and R. Maynard, Phys. Rev. Lett. $\bf 56$ (1986), 1471-1474

\bibitem{Akk288} E. Akkermans, P.E. Wolf, R. Maynard and G. Maret,
J. de Physique (France), $\bf 49$ (1988), 77-98

\bibitem{vdh} H. C. van de Hulst, {\it Multiple light scattering} (Dover, N.Y. 1981).

\bibitem{Bohigas91}O. Bohigas, in {\it Chaos and Quantum Physics,
Proceedings of
the  Les Houches Summer School, Session LII}, ed. by M.J. Giannoni, A. Voros
and J. Zinn-Justin (North Holland, Amsterdam, 1991), p.91

\bibitem{Althuler86}B.L. Al'tshuler and B. Shklovski\u{\i}, Sov. Phys. JETP {\bf 64}, 127 (1986).
\end{thebibliography}
\end{document}